# Analysis of an Evaporating Sessile Droplet on a Non-Wetted Surface

Rajneesh Bhardwaj
Department of Mechanical Engineering,
Indian Institute of Technology Bombay,
Mumbai 400076, India
Email: rajneesh.bhardwaj@iitb.ac.in

*Abstract*

We investigate evaporation of a sessile droplet on a non-wetted surface in the framework of diffusion-limited and quasi-steady evaporation. We extend previous models and numerically solve Laplace equation for the diffusion of liquid vapor in ambient. We propose a unified, simple and accurate expression of the evaporation mass flux valid for 90° ≤ $\theta$ ≤ 180°, where $\theta$ is the equilibrium contact angle. In addition, using the derived expression of the evaporation mass flux, we propose a simple and accurate expression of the evaporation mass rate for a non-wetted surface, which does not exhibit singularity at $\theta$ = 180°. Finally, using the scaling analysis, the expression of the evaporation mass flux is utilized to estimate the direction and magnitude of the characteristic evaporation-driven flow velocity inside the droplet on a non-wetted surface. The predicted flow direction is found to be consistent with the previous measurements.

*Keywords*: Evaporating sessile droplet, Non-wetted surface, Evaporation mass flux, Evaporation mass rate

---

Owing to several technical applications, the evaporation of a sessile droplet on a solid surface is a much-studied problem in the interface science in the last decade [1]. In particular, an evaporating droplet can be utilized to self-assemble colloidal particles suspended in it [2]. In the framework of quasi-steady and diffusion-limited evaporation, previous studies [2, 3] have shown that the evaporation mass flux ($J$ [kg m$^{-2}$ s]) on the liquid-gas interface is non-uniform on a partially-wetted substrate (0° < $\theta$ ≤ 90°) and the largest evaporation occurs near the contact line. Hu and Larson [3] simplified Deegan's model [2] and provided the following simplified expression of $J$ for a partially-wetted surface,

$$J(r) = J_0(\theta)\left[1-\left(\frac{r}{R}\right)^2\right]^{-\lambda(\theta)}, \qquad (1)$$

where $r$, $\theta$ and $R$ are the radial coordinate, contact angle and wetted radius respectively. The expressions of $J_0(\theta)$ and $\lambda(\theta)$ are given as follows [3],

$$\begin{aligned} J_0(\theta) &= [D(c_{sat}-c_\infty)/R](0.27\theta^2+1.30)(0.6381-0.2239(\theta-\pi/4)^2), \\ \lambda(\theta) &= 0.5-\theta/\pi \end{aligned} \qquad (2)$$

where $D$ is the diffusion coefficient [m$^2$ s$^{-1}$] of the liquid vapor in outside gas, $c_{\mathrm{sat}}$ is vapor concentration [kg m$^{-3}$] at its saturated value at the ambient temperature, $c_\infty$ is the concentration value in the ambient and $R$ is the wetted radius of the droplet. The evaporation mass flux diverges at the contact line ($r = R$) for $\theta < 90°$ [2, 3] and a constant value for $\theta = 90°$ is given by the following expression [4],

$$J = \frac{D(c_{sat} - c_\infty)}{R} \tag{3}$$

The largest error between the numerical solution as compared to the fitted solution obtained by eq. 1 was around 6% [3].

In case of a non-wetted surface ($90° < \theta \leq 180°$), the largest evaporation occurs at the apex of the droplet [5]. The non-wettability or a larger contact angle can be achieved by engineering nano- and micro-textures on a surface with contact angle larger than 65° [6, 7]. Popov [4] derived the following generalized expression of $J$, valid for any contact angle, $0° < \theta \leq 180°$.

$$J = \frac{D(c_{sat} - c_\infty)}{R}\left[\frac{1}{2}\sin\theta + \sqrt{2}(\cosh\alpha + \cos\theta)^{3/2}\int_0^\infty \frac{\tau\cosh\theta\tau}{\cosh\pi\tau}\tanh\left[\tau(\pi - \theta)\right]P_{-1/2+i\tau}(\cosh\alpha)\, d\tau\right] \tag{4}$$

where $\alpha$ and $P_{-1/2+i\tau}$ are toroidal coordinate and Legendre functions of the first kind, respectively [4]. Stauber et al. [5] revisited solution of an equivalent electrostatics problem reported by Smith and Barakat [8] and derived the following closed form of $J$ for a non-wetted surface at $\theta = 180°$,

$$J(z) = \frac{D(c_{sat} - c_\infty)}{2R_{sph}}\left[1 + \left(\frac{2R_{sph}}{z}\right)^{3/2}\int_0^\infty q\tanh B_0\left(\frac{rq}{z}\right)\exp(-q)\ \mathrm{dq}\right] \tag{5}$$

where $B_0(\cdot)$ is the Bessel function of the first kind of zeroth order, $R_{\mathrm{sph}}$ is the radius of the sphere fitted to the droplet ($R_{\mathrm{sph}} = R/\sin\theta$) and $z$ is the axial coordinate on the liquid-gas interface, expressed as [5], $z = -R_{sph}\cos\theta \pm \sqrt{R_{sph}^2 - r^2}$ , where + and – sign corresponds to the upper and lower half of the sphere, respectively. Authors of recent studies [5, 9, 10] plotted the solution of eq. 4 and showed that the profile of $J$ for a non-wetted surface is significantly different than that for a partially-wetted surface [2, 3]. Specifically, in the former, $J$ slightly decreases along the upper half of the droplet and decays to a smaller value near the contact line in the lower half.

The expression of $J$ for non-wetted surface ($90° < \theta \leq 180°$) reported in the literature (eq. 4) is transcendental and it is not easy to use it in the simple models. A simplified expression of the evaporation mass flux (eq. 1) was reported by Hu and Larson [3] for a partially-wetted surface (0° < $\theta \leq$ 90°). However, to the best of our knowledge, an expression for a non-wetted surface has not been reported in the literature. Similarly, the expression of evaporation mass rate $\dot{M}$ [kg $s^{-1}$], reported by Hu and Larson [3], is valid for 0° ≤ $\theta \leq$ 90° and a simple expression of $\dot{M}$, recently reported by Hu et al. [11], exhibits singularity at $\theta$ = 180°. In addition, using scaling analysis, previous studies have shown the dependence of the magnitude of the internal evaporation-driven flow velocity on $J$ on the partially-wetted surface. For instance, the internal flow velocity scales with $J$ near the contact line in the absence of Marangoni stresses [12]. The scaling analysis of the internal flow velocity on a non-wetted surface has not been reported thus far, to the best of our knowledge. Therefore, the objective of this letter is to derive simple and accurate expressions of the evaporation mass flux and evaporation rate of an evaporating sessile droplet on a non-wetted surface (90° < $\theta \leq$ 180°). A secondary objective is to estimate the direction and magnitude of the internal velocity by scaling analysis, using the derived expression of $J$.

First, we numerically integrate eq. 5 to obtain $J$ at $\theta$ = 180° and compare our data with solution of an equivalent electrostatics problem, reported by Smith and Barakat [8]. In this problem, the electrostatic potential field is solved around two perfectly conducting contiguous spheres of the same radius. The vapor concentration and evaporation flux correspond to the electrostatic potential and surface charge density, respectively. Fig. 1 shows a good agreement of the variation of normalized evaporation flux ($J_N = JH/[D(c_{sat} - c_\infty)]$, where $H = 2R_{sph}$ is droplet height) with respect to azimuthal angle $\phi$, obtained in present work and that reported by Smith and Barakat [8]. The direction of $\phi$ is shown in the inset, and $\phi$ = 0° and $\phi$ = 180° correspond to the apex of the droplet and to the contact line, respectively. The evaporation flux on the upper hemisphere slightly decreases and the value at $\phi$ = 90° decreases by 12% of the value at $\phi$ = 0°. In the lower half of the hemisphere, $J$ decays exponentially to zero at $\phi$ = 180°.

Second, a finite element method based numerical model is employed for simulating the diffusion-limited and quasi-steady evaporation of a sessile, spherical cap droplet on a non-wetted surface at ambient temperature (Fig. 2(a)). We solve the diffusion of the liquid vapor in the air surrounding the droplet using Laplace equation [3], $\vec{\nabla}^2 c = 0$, where $c$ is the liquid vapor concentration [kg m$^{-3}$]. The evaporation mass flux at the liquid-gas interface ($J$) is expressed as follows, $\vec{J} = -D\vec{\nabla}c\big|_{LG}$, where subscript LG denotes the liquid-gas interface. We solve the Laplace equation in a computational domain shown in Fig. 2(b). The boundary conditions are shown in Fig. 2(b) and are briefly described as follows. The vapor concentration at the droplet-air interface is prescribed at its saturated value at the ambient temperature ($T_\infty$ = 25°C), $c$ = $c_{sat}$. The concentration in the far-field is expressed in term of relative humidity of the ambient ($\gamma$) and is given by $c = c_\infty = \gamma c_{sat}$. The value of $\gamma$ is taken as 0.5 in the simulations. The far-field is set at $r$ = 50$R$, $z$ = 50$R$, where $R$ is the wetted radius of the droplet, after performing a domain-size independence study. Axisymmetric boundary condition, $\partial c/\partial r = 0$, is applied at, $z > H$, $r = 0$, where $H$ is the height of the droplet. No penetration of the vapor concentration into the surface of the substrate, $\partial c/\partial z = 0$, is applied at $r > R$, $z = 0$. The following parameters are used in the model [3]: $D = 2.61 \times 10^{-5}$ m$^2$/s and $c_{sat} = 2.32 \times 10^{-3}$ kg/m$^3$. A grid-size convergence study is performed to select adequate grid resolution and a typical grid used in the simulations is shown in Fig. 2(c). The validations of the model are included in the supplementary data.

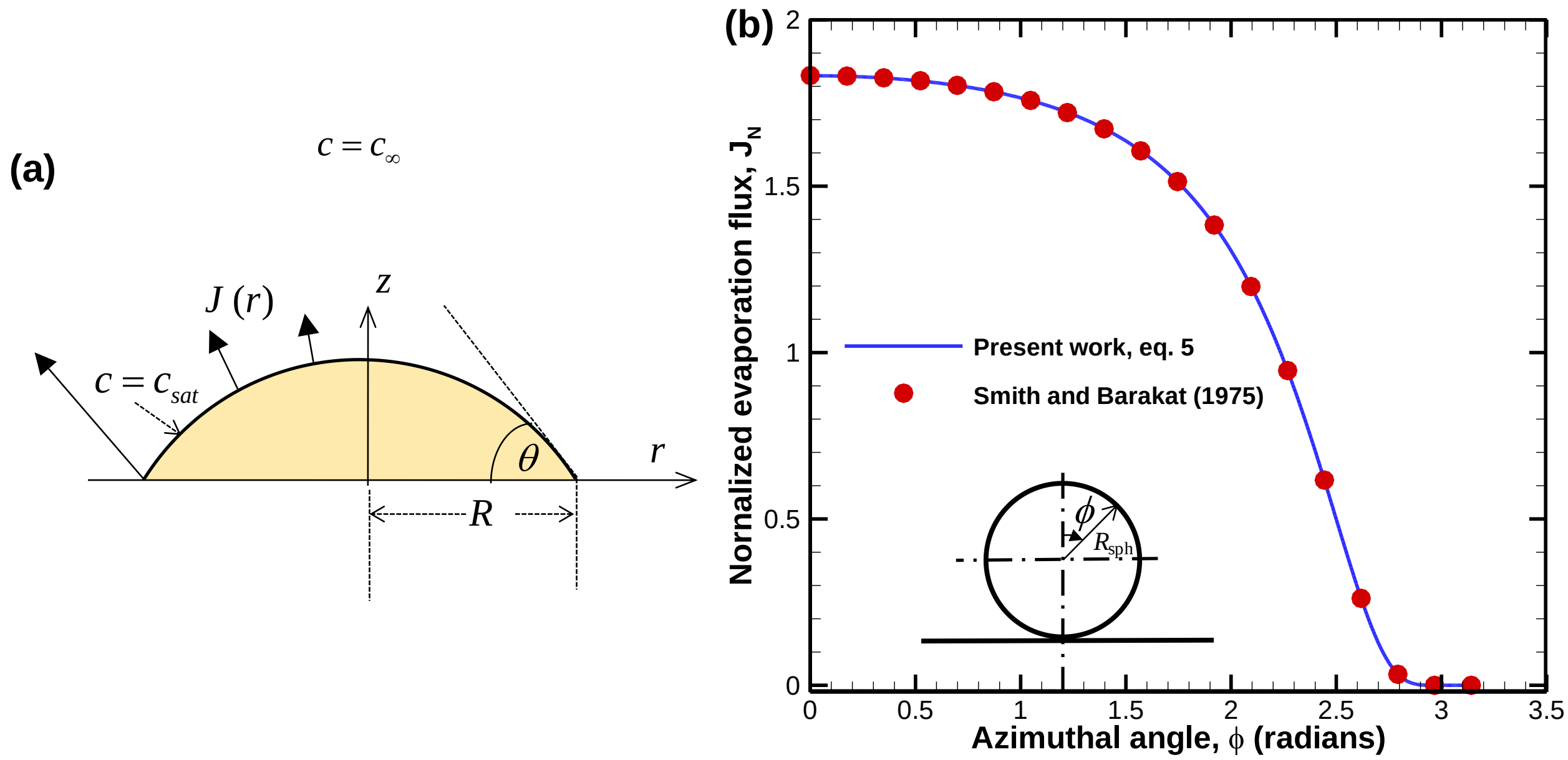

Fig. 1: (a) Geometry of a sessile droplet on a partially-wetted surface (b) Comparison between computed normalized evaporation flux $J_N$ using eq. 5 for $\theta = 180^{\circ}$ in the present work with that reported by Smith and Barakat [8] for equivalent electrostatics problem. The inset shows that geometry of the droplet considered in this case.

In the limiting case of $\theta = 180^{\circ}$, the coordinates of the apex of the droplet are, $r = 0$, $z = H = 2R_{\text{sph}}$, $h_N = 1$, where $h_N$ is normalized axial coordinate ($h_{\text{N}} = h/H$). The normalized flux is, $J_{\text{N}} = 2C$, where $C$ is Catalan constant ($C$= 0.916, $2C - 1 = \int_0^{\infty} q \tanh q \, \mathrm{e}^{-q} \, \mathrm{d}q$), given by eq. 5 because $B_0(0) = 1$ in this case. The flux, $J_{\text{N}}$, at the contact line ($h_N = 0$, i.e. $\phi = 3.14$) is $J_{\text{N}} = 0$, as plotted in Fig. 1. Owing to exponential decay of $J_{\text{N}}$ with respect to $h_N$ (Fig. 1), a simplified expression of $J_{\text{N}}$ for $\theta = 180^{\circ}$ is proposed as follows,

$$J_N(h_N, \theta) = J_0(\theta)\left[1 - e^{-b(\theta) h_N}\right]^{-\lambda(\theta)} \qquad (6)$$

where $J_0$, $b$ and $\lambda$ are functions of $\theta$, and eq. 6 satisfies $J_{\text{N}} = 0$ at $h_N = 0$. Using curve fitting by least squares method, $J_0$, $b$ and $\lambda$ are obtained as $2C$, 5.503 and -1.5, respectively, with $R^2 = 0.998$. In the limiting case of $\theta = 90^{\circ}$, $J_{\text{N}}$ is constant and equal to 1 (eq. 3), and thereby, $J_0$ and $\lambda$ are 1 and 0, respectively, in order to extend eq. 6 to this case.

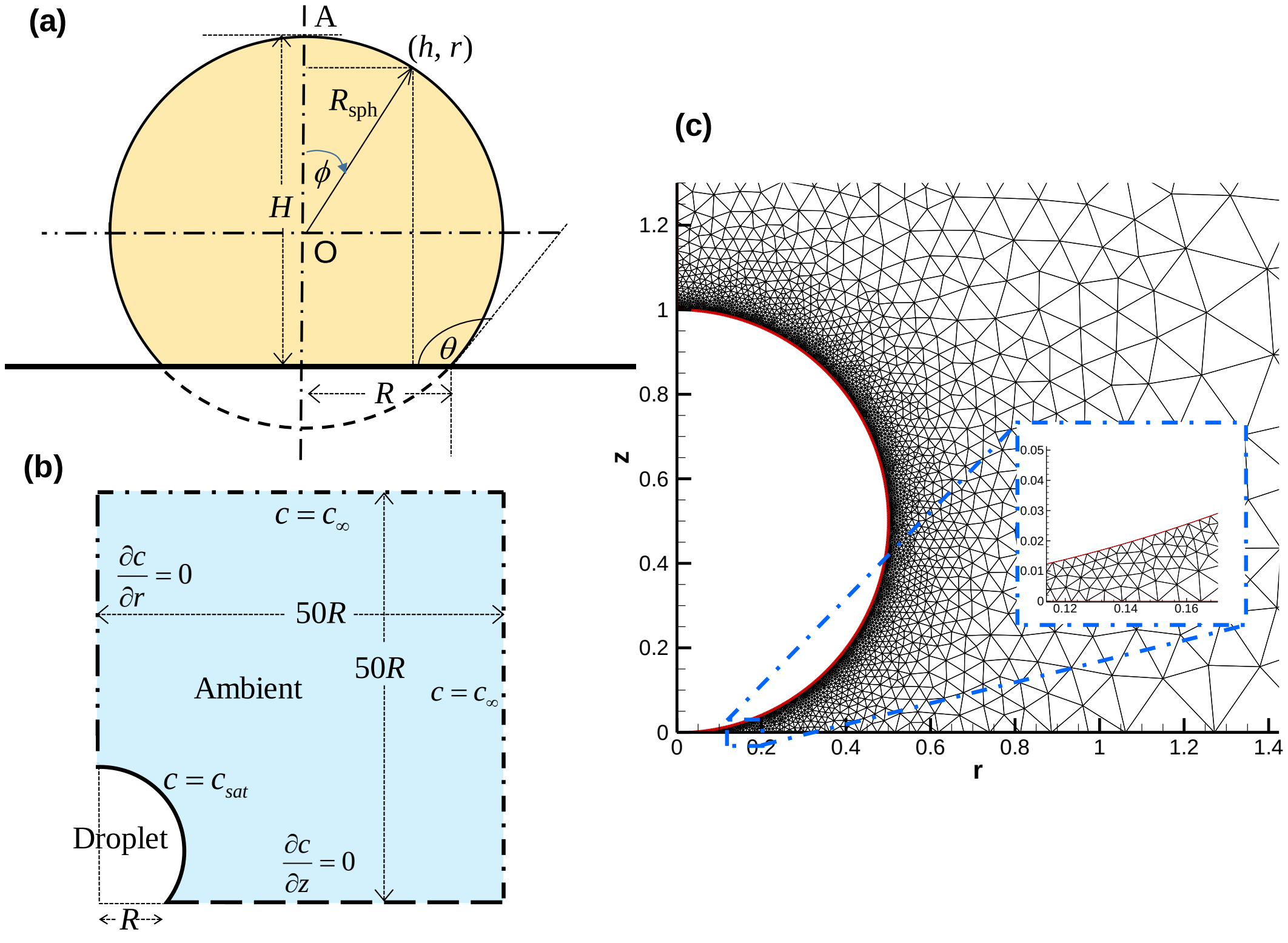

Fig. 2: (a) Geometry of a sessile droplet on a non-wetted surface (b) Computational domain (not to scale) shown as a shaded area and boundary conditions used in numerical simulations (c) A typical finite element

mesh used in the simulations for $\theta = 177°$. The inset shows a zoomed-in view of the mesh near the contact line outside the droplet.

For $90^o < \theta < 180^o$, $J_N$ is not zero at $h_N = 0$ (at the contact line) and eq. 6 cannot be satisfied at $h_N = 0$ for $90^o < \theta < 180^o$. In order to extend eq. 6 for $90^o < \theta < 180^o$ and to alleviate this problem, eq. 6 is slightly modified as,

$$J_N(h_N,\theta) = J_0(\theta)\left[k(\theta) - e^{-b(\theta)h_N}\right]^{-\lambda(\theta)} \tag{7}$$

where $J_0$, $k$, $b$ and $\lambda$ are functions of $\theta$. Note that eq. 7 is satisfied for $\theta = 90^o$ with $k = 1$, $J_0 = 1$ and $\lambda = 0$. Using the finite-element model, we performed simulations at different contact angles and contours of the vapor concentration are plotted in Fig. 3(a-c). The value of the vapor concentration near the contact line at $\theta = 177^o$ in Fig. 3(c) is equal to that on the liquid-gas interface and hence it qualitatively confirms almost zero value of the evaporation flux near the contact line in this case. We fit exponential curves, described by eq. 7, to the numerical profiles of $J_N$ for different values of $\theta$, using least squares method. The values of the fitting constants with $R^2$ values at different contact angles are listed in Table S1 in the supplementary data. The $R^2$ values for all fits are closer to 0.99. Comparisons between the simulated and fitted profiles for $90^o \leq \theta \leq 180^o$ in Fig. 3(d) show a good agreement. The maximum difference between the two profiles with respect to the computed value is around 3%. In addition, the fitted values of $J_N$ at $h_N = 0$ are very close to the computed ones (given in Table S2 in the supplementary data). $J_0$, $k$, $b$ and $\lambda$ in eq. 7 are obtained as polynomial fits using data in Table S1 and are expressed as follows,

$$\begin{aligned} J_0(\theta) &= -0.29\theta^2 + 1.94\theta - 1.34, \\ k &= 1 + 31.86e^{-3.54\theta}, \\ b &= 0.61\theta^2 - 2.12\theta + 6.16, \\ \lambda &= 3(0.5 - \theta/\pi) \end{aligned} \tag{8}$$

The $R^2$ values of the fittings in eq. 8 are larger than 0.97 (listed in Table S3 in the supplementary data). In eq. 8, the profile of $\lambda$ with respect to $\theta$ is linear, similar to that obtained by Hu and Larson [3] for $0^o < \theta < 90^o$. Note that $\lambda = 0$ corresponds to $\theta = 90^o$ in both studies. However, $\lambda \geq 0$ is valid for $0° < \theta \leq 90°$, as reported by Hu and Larson [3] while $\lambda \leq 0$ is obtained in the present work for $90^o \leq \theta \leq 180^o$. This is due to the fact that the evaporation flux decays to zero near the contact line for the non-wetted surface in the present work while it is the largest near the contact line for the partially-wetted surface. Eqs. 7 and 8, therefore, represent the simplified expression of the evaporation mass flux valid for $90^{\circ} \leq \theta \leq 180^{\circ}$.

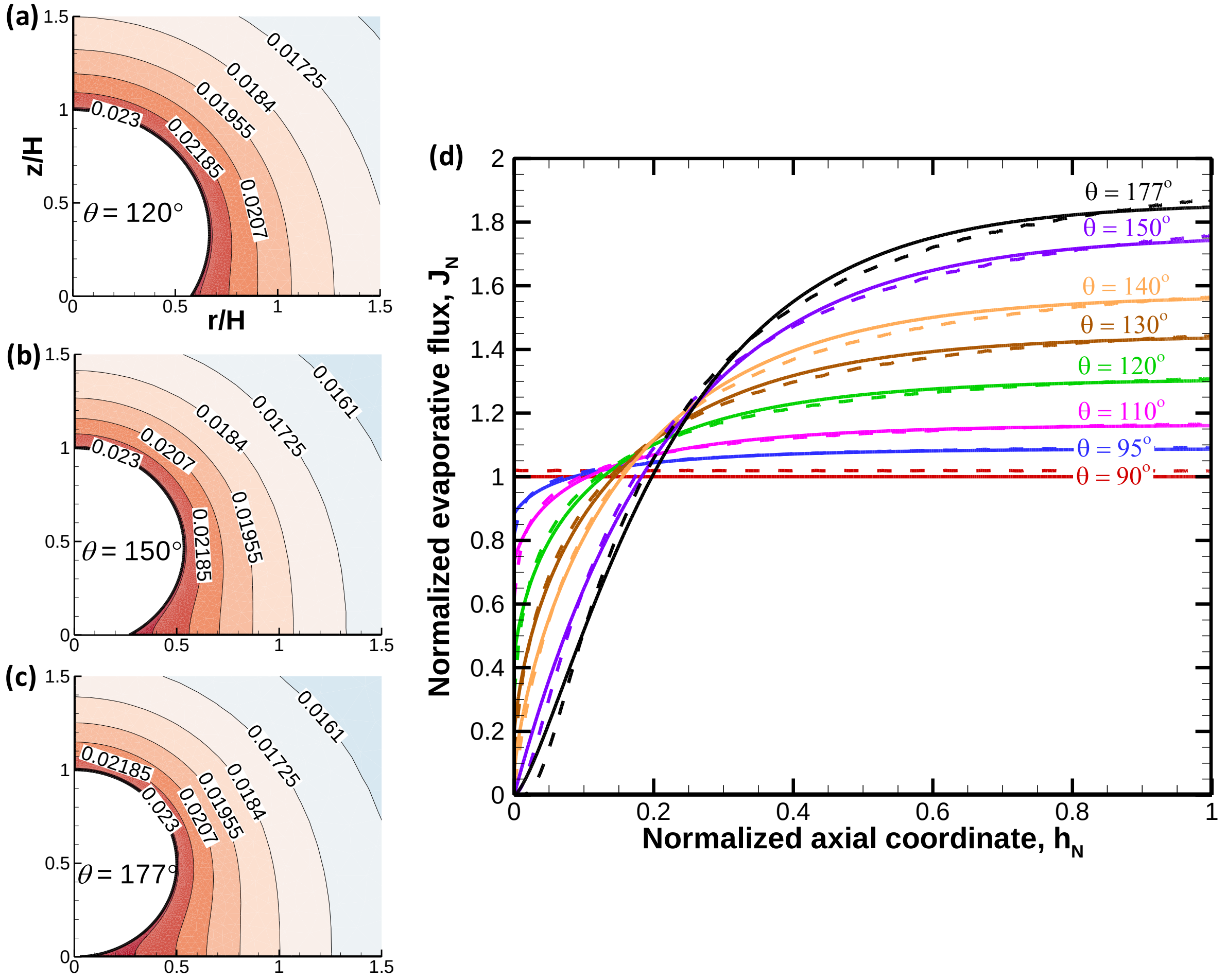

Fig. 3: (a) Contours of vapor concentration for different cases of contact angles in the range of $90^{o} \leq \theta < 180^{o}$. (b) Fitted profiles of the evaporation flux at different contact angles for $90^{o} \leq \theta < 180^{o}$ as function of normalized axis coordinate ($h_N = h/H$).

Third, we calculate the normalized evaporation mass rate ($\dot{M}_N$) based on droplet height ($H$) by integrating $J$ (eq. 7) over the liquid-gas interface and is given as follows,

$$\dot{M}_N = \frac{\dot{M}}{HD(c_{sat} - c_{\infty})} = -\frac{1}{HD(c_{sat} - c_{\infty})}\int_A \vec{J}(h,\theta)\cdot\vec{n}ds = -\frac{1}{HD(c_{sat} - c_{\infty})}\int_A J(h,\theta)\left(\sqrt{\left(\frac{\partial h}{\partial r}\right)^2 + 1}\right)2\pi r dr \quad (9)$$

The expression of $\dot{M}$ valid for any contact angle was reported by Popov [4] and after normalizing $\dot{M}$ with droplet height ($H$), we obtain $\dot{M}_N$ as follows,

$$\dot{M}_N = \frac{\dot{M}}{HD(c_{sat} - c_\infty)} = -\frac{\pi}{\tan(\theta/2)}\left[\frac{\sin\theta}{1+\cos\theta} + 4\int_0^\infty \frac{1+\cosh 2\theta\tau}{\sinh 2\pi\tau}\tanh\left[(\pi-\theta)\tau\right]d\tau\right] \qquad (10)$$

We numerically solve eqs. 9 and 10 for different values of $\theta$ and compare profiles of $\dot{M}_N$ in Fig. 4. The values obtained by eq. 9 are in good agreement with those obtained by eq. 10, with a maximum difference of 3%. Therefore, the comparison verifies the fidelity of the proposed expression of $J$ (eqs. 7-8). The profile obtained by eq. 9 is almost overlapping with a series solution proposed by Picknett and Bexon [13] and is also closer to a profile recently reported by Hu et al. [11]. In Fig. 4, the computed values obtained using eq. 9, $\dot{M}_N \approx 2\pi$ at $\theta = 90°$ and $\dot{M}_N \approx 1.39\pi$ at $\theta = 180°$ are consistent with the studies of Hu and Larson [3], and Picknett and Bexon [13], respectively. In this context, $\dot{M}_N$ for a suspended droplet is $2\pi$ (as derived in the supplementary data) and is expected to be same as obtained for $\theta = 90°$. For $\theta = 180°$, $\dot{M}_N \approx 1.39\pi$ is equivalent to that for a suspended droplet touching an infinite wall, which suppresses the diffusion of liquid vapor and reduces $\dot{M}_N$ to $1.39\pi$. Further, we propose the following simple and accurate profile of $\dot{M}_N$, that is valid for $90^o \leq \theta < 180^o$ and is obtained by the curve fitting to the solution of eq. 10, using least squares method ($R^2 = 0.999$),

$$\dot{M}_N = 6.11\exp(-0.45\theta^2) + 4.28 \qquad (11)$$

The profile obtained by eq. 11 is plotted in Fig. 4 and almost overlaps with that obtained by eq. 10. Note that there is a singularity in the limit of $\theta \to 180°$ in the solution of eq. 10, also reported by Hu et al. [11]. However, eq. 11 does not exhibit the singularity at $\theta \to 180°$.

Finally, we propose an expression for the internal evaporation-driven velocity on a non-wetted surface using the expression of $J$ derived earlier. As explained by Deegan [2], in case of $0^o < \theta \leq 90^o$, the maximum evaporation near the contact line together with the pinning of the contact line results in a radially outward evaporation-driven flow. If the droplet is loaded with colloidal particles, they advect towards the contact line to form a ring-like deposit [2], as shown in a schematic in Fig. 5(a). Based on the expression of the radial velocity reported by Hu and Larson [14], a scaling of internal evaporation-driven velocity, $V_{\mathrm{rad}}$, was reported in Ref. [12] and is given by, $V_{rad} \sim J_{\max} / \rho$, where $J_{\max}$ is the maximum evaporation occurs near the contact line and $\rho$ is the density of the liquid.

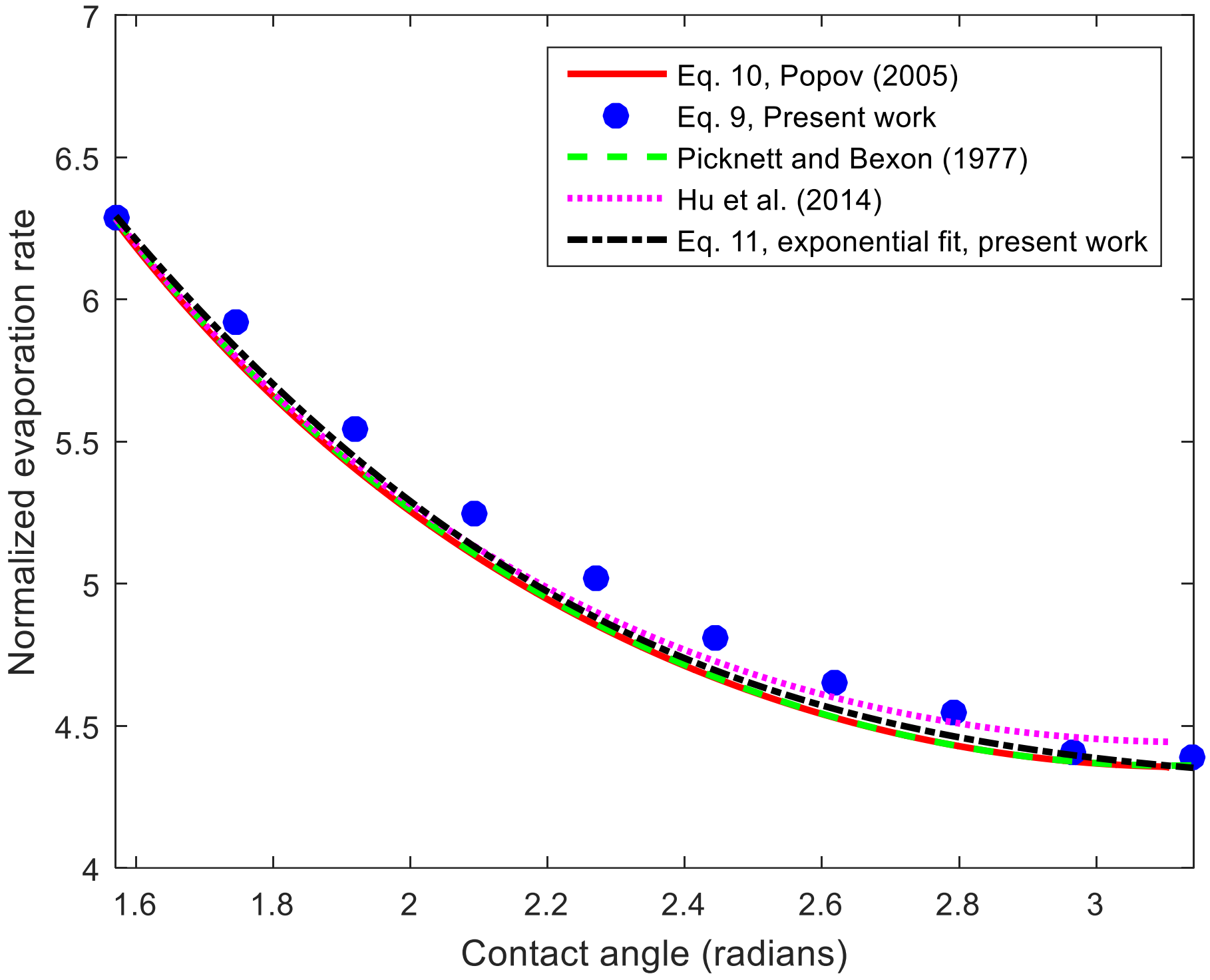

Fig. 4: Comparison of variation of the normalized evaporation mass rate with contact angle obtained by numerical solution of eq. 10, and by numerically integrating evaporation flux over the liquid-gas interface (eq. 9). The published profiles are also compared. A proposed exponential fit (eq. 11) is plotted and almost overlaps with the computed solution of eq. 10.

In the present case of a non-wetted surface, $90^{\circ} \leq \theta \leq 180^{\circ}$, the maximum evaporation flux occurs near the apex of the droplet (Fig. 3(d)) and is given by, $J_{\max} = J_0(\theta)D(c_{\mathrm{sat}} - c_{\infty})/H$ (eq. 7). Therefore, a key difference between the evaporation on the partially-wetted ($0^{\circ} < \theta \leq 90^{\circ}$) and non-wetted ($90^{\circ} < \theta \leq 180^{\circ}$) surface is the location of the maximum value of $J$ on the liquid-gas interface. In the former, it is near the contact line, while in the latter, it is at the apex of the droplet. Since $J$ exponentially decays to zero near the contact line for $\theta \rightarrow 180^{\circ}$ for the latter, we postulate a flow direction from the edge of the droplet to the center of the droplet, as shown in a schematic in Fig. 5(b) for a non-wetted surface ($90^o < \theta \leq 180^o$). The axial velocity $V_{axi}$ scales as, $V_{axi} \sim J_{max} / \rho \sim J_0(\theta) D(c_{sat} - c_\infty)/(\rho H)$. At $\theta \geq 150^o$, $J_0(\theta)$ is approximated as $2C$ within 10% error (Fig. 3(d)) and $V_{axi}$ scales as, $V_{axi} \sim CD(c_{sat} - c_\infty)/(\rho R)$, where $C$ is Catalan constant and $R$ is the wetted radius. Reported experiments provide corroboration of the proposed postulation [15, 16]. For instance, Manukyan et al. [15] imaged internal velocity profile in an evaporating droplet on a hydrophobic surface and showed that a convection current occurs from the contact line to the center of the droplet for a non-wetted surface. Similarly, Chen and Evans [16] showed that the colloidal particles move upward due to larger mass loss at the apex of the droplet as compared to near the contact line.

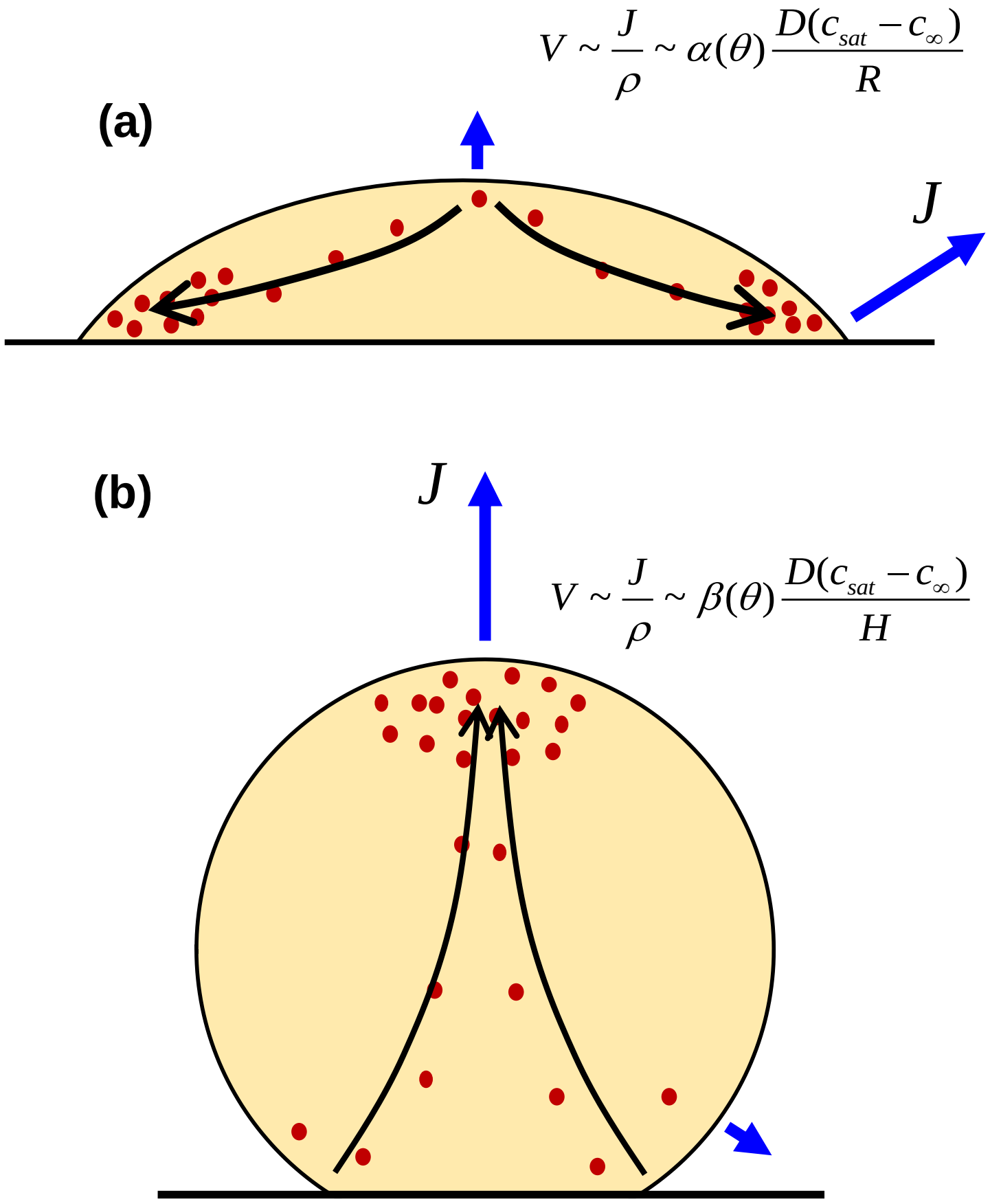

Fig. 5: Schematic illustration of the advection of colloidal particles (shown as red dots) by evaporation driven flow on partially wetted (a) and non-wetted (b) surface. Blue arrows compare the magnitude of the evaporation flux near the contact and at the apex of the droplet. The respective velocity scaling is also shown.

In closure, we have numerically investigated diffusion-limited and quasi-steady evaporation of a sessile droplet on a non-wetted surface. We have used a finite element numerical model to solve Laplace equation for the diffusion of water vapor into the air surrounding the droplet. The model does not account for Marangoni stresses, substrate heating and buoyancy-induced convection. The numerical results are validated against available benchmark data of the evaporation mass flux. Based on the numerical results, we propose simple and accurate expressions of evaporation mass flux ($J$) and evaporation mass rate ( $\dot{M}$ ). The maximum difference between the simulated and proposed expression of $J$ is found to be lesser than 3%.The expression of $\dot{M}$ does not exhibit singularity at $\theta = 180^{\circ}$. We obtain the magnitude of the internal velocity using scaling analysis and expression of $J$. The direction of the velocity on a non-wetted surface is found to be opposite to that on the partially-wetted surface and is consistent with the previous measurements. The present work provides fundamental insights into the variation of $J$ with space and contact angle, and variation of $\dot{M}$ with contact angle for a non-wetted surface. The proposed simple and accurate expressions of $J$ and $\dot{M}$ will be useful to build complex models for addressing coupled mass, momentum and heat transport during the evaporation.

## Acknowledgments

R.B. gratefully acknowledges financial support of a CSR grant from Portescap Inc., India and of an internal grant from Industrial Research and Consultancy Centre (IRCC), IIT Bombay.

## References

[1] D. Brutin, *Droplet Wetting and Evaporation: From Pure to Complex Fluids*. Tokyo: Academic press, 2015.

[2] R. D. Deegan, O. Bakajin, T. F. Dupont, G. Huber, S. R. Nagel, and T. A. Witten, "Capillary flow as the cause of ring stains from dried liquid drops," *Nature,* vol. 389, pp. 827-829, 1997.

[3] H. Hu and R. G. Larson, "Evaporation of a sessile droplet on a substrate," *J. Phys. Chem. B* vol. 106, pp. 1334-1344, 2002.

[4] Y. O. Popov, "Evaporative deposition patterns: Spatial dimensions of the deposit," *Phys. ReV. E,* vol. 71, pp. 0363131- 03631317, 2005.

[5] J. M. Stauber, S. K. Wilson, B. R. Duffy, and K. Sefiane, "Evaporation of droplets on strongly hydrophobic surfaces," *Langmuir,* vol. 31, p. 3653, 2015.

[6] Y. Tian and L. Jiang, "Wetting: Intrinsically robust hydrophobicity," *Nature materials,* vol. 12, pp. 291-292, 2013.

[7] Y. Tian, B. Su, and L. Jiang, "Interfacial material system exhibiting superwettability," *Advanced Materials,* vol. 26, pp. 6872-6897, 2014.
[8] G. S. Smith and R. Barakat, "Electrostatics of two conduction spheres in contact," *Appl. Sci. Res. ,* vol. 30, p. 418, 1975.
[9] T. A. Nguyen and A. V. Nguyen, "On the lifetime of evaporating sessile droplets," *Langmuir,* vol. 28, pp. 1924-1930, 2012.
[10] S. Dash and S. V. Garimella, "Droplet Evaporation Dynamics on a Superhydrophobic Surface with Negligible Hysteresis," *Langmuir,* vol. 29, p. 10785−10795, 2013.
[11] D. Hu, H. Wu, and Z. Liu, "Effect of liquid-vapor interface area on the evaporation rate of small sessile droplets," *International Journal of Thermal Sciences,* vol. 84, pp. 300-308, 2014.
[12] R. Bhardwaj, X. Fang, P. Somasundaran, and D. Attinger, "Self-assembly of colloidal particles from evaporating droplets: role of DLVO interactions and proposition of a phase diagram," *Langmuir,* vol. 26, pp. 7833-7842, 2010.
[13] R. G. Picknett and R. Bexon, "The evaporation of sessile or pendant drops in still air," *J. Colloid Interface Sci.,* vol. 61, pp. 336-350, 1977.
[14] H. Hu and R. G. Larson, "Analysis of Microfluid flow in an Evaporating sessile droplet," *Langmuir,* vol. 21, pp. 3963-3971, 2005.
[15] S. Manukyan, H. M. Sauer, I. V. Roisman, K. A. Baldwin, D. J. Fairhurst, H. Liang*, et al.*, "Imaging internal flows in a drying sessile polymer dispersion drop using Spectral Radar Optical Coherence Tomography (SR-OCT)," *Journal of Colloid and Interface Science,* vol. 395, pp. 287-293, 2013.
[16] L. Chen and J. R. G. Evans, "Drying of colloidal droplets on superhydrophobic surfaces," *Journal of Colloid and Interface Science,* vol. 351, pp. 283-287, 2010.